\documentclass{emulateapj}
\newcommand{\revhk}[1]{#1}

\begin{document}

\title{From Planetesimal to Planet in Turbulent Disks. II. Formation of
Gas Giant Planets}

\shorttitle{
Formation of Gas Giant Planets
}

\author{Hiroshi Kobayashi}
\affil{
Department of Physics, Nagoya University, Nagoya, Aichi 464-8602, Japan
}
\email{hkobayas@nagoya-u.jp}

\author{Hidekazu Tanaka}
\affil{
Astronomical Institute, Tohoku University, Aramaki, Aoba-ku, Sendai
980-8578, Japan
}
\email{hidekazu@astr.tohoku.ac.jp}

\begin{abstract}
In the core accretion scenario, 
gas giant planets are formed form 
solid cores with several Earth masses via gas accretion. 
We investigate the formation of such cores 
via collisional growth from kilometer-sized planetesimals 
in turbulent disks. 
The stirring by forming cores induces collisional fragmentation and
 surrounding planetesimals are ground down until radial drift. 
 The core growth is therefore stalled by the depletion of surrounding
 planetesiamls due to collisional fragmentation and radial drift. 
The collisional strength of planetesimals determines 
the planetesimal-depletion timescale, 
which is prolonged for large planetesiamls. 
The size of planetesiamls around growing cores is determined by the planetesimal size distribution at the onset of runaway growth. Strong turbulence delays 
the onset of runaway growth, resulting in large planetesimals. 
Therefore, the core mass evolution depends on turbulent parameter $\alpha$; 
the formation of cores massive enough 
without significant depletion of surrounding planetesimals needs strong
 turbulence of $\alpha \ga 10^{-3}$. 
However, the strong turbulence with $\alpha \ga 10^{-3}$ leads to a
 significant delay of the onset of runaway growth and prevents the formation of massive cores
 within the disk lifetime. 
The formation of cores massive enough 
within several millions years therefore requires the several times enhancement of the solid surface
 densities, which is achieved in the inner disk $\la 10$AU due to pile-up of drifting
 dust aggregates. In addition, the collisional strength $Q_{\rm D}^*$
 even for kilometer-sized or smaller bodies
 affects the growth of cores; $Q_{\rm D}^* \ga 10^7 \,{\rm erg/g}$ for
 bodies $\la 1$\,km is likely for this gas giant formation. 

\end{abstract}

\section{Introduction}

Gas giant planets such as Jupiter, Saturn, and massive exoplanets are
formed in protoplanetary disks containing solid and gas. In the core
accretion scenario, once planetary embryos grow to $\sim 5 M_\oplus$, 
planetary atmospheres that Mars sized or
larger embryos acquire become too massive to keep hydrostatic states,
resulting in the rapid gas accretion to form gas giants \citep[e.g.,][]{mizuno80}. 
Therefore, gas giant formation requires such massive solid cores to be formed
within the disk lifetime ($\sim 4\,$Myr). 

In protoplanetary disks, sub-micron sized dust grains accumulate to be
planetesimals. Collisional growth of grains produces centimeter sized or
larger particles, which drift onto the host star on significantly short timescales 
due to moderate coupling
with gas. If particles accumulate up to the Roche density, gravitational
instability creates kilometer sized or larger planetesimals
\citep{goldreich,youdin,michikoshi, takeuchi}. On the other hand, collisional
growth of dust grains naturally form fluffy aggregates \citep{suyama08,suyama12}. 
The bulk density of dust aggregates becomes $\sim 10^{-4} \,
{\rm g/cm}^3$ and the collisional growth timescale of such bodies is
much shorter than  their drift timescale, so that kilometer sized planetesimals are formed via
direct collisional growth prior to radial drift \citep{okuzumi12}. 

Planetesimals grow through mutual collisions. In a turbulent disk, 
the stirring by turbulence increases the random velocity of
planetesimals $v_{\rm r}$. 
For large $v_{\rm r}$, the gravitational focusing of planetesimals is negligible and the
orderly growth occurs until $v_{\rm r} \ga 1.5 v_{\rm esc}$ \citep{kobayashi16}, where
$v_{\rm esc}$ are the surface escape velocities of planetesimals. In the orderly growth, the mass-weighted average
radius of planetesimals is comparable to the radius of largest planetesimals. As planetesimals grow,
$v_{\rm esc}$ increases to $\sim v_{\rm r}$, runaway growth occurs
\citep{wetherill89}.  
The mass-weighted average radius of planetesimals at the onset of
runaway growth, which is determined by turbulent strength
\citep{kobayashi16}, almost remains the same after runaway growth. 

Runaway growth forms a planetary embryo in each annulus of the
disk. 
Embryos further grow mainly through collisions with surrounding planetesimals. 
As embryos become massive, their viscous stirring increases $v_{\rm r}$,
resulting in destructive collisions between planetesimals. 
Collisional fragments of planetesimals get further smaller via
collisions between themselves. This collisional cascade decreases the
size of bodies until radial drift. This process reduces surrounding
planetesimals and embryo growth is then stalled \citep{kobayashi+10}. 

Collisional strength depends on the radius of planetesimals, $r$
\citep[e.g.,][]{benz99}. For $r \ga 1$\,km, the collisional outcome of a
single collision is controlled by the self-gravity of colliders so that
larger bodies are effectively stronger for collisions. 
Larger planetesimals, which have a longer collisional depletion timescales,
contribute more to the growth of massive embryos. Embryo growth strongly
depends on the mass-weighted average radius of planetesimals surrounding
embryos \citep{kobayashi+10,kobayashi11,kobayashi13}, which is mainly determined
by the strength of turbulence \citep{kobayashi16}. Therefore, the growth and
formation of solid cores of gas giants depends on the strength of disk
turbulence. 

In this paper, we investigate gas giant planet formation via core
accretion in turbulent disks. In \S~\ref{sc:ccm}, we introduce the
critical core mass from simple analysis. In \S~\ref{Sc:model}, 
we explain the model of simulations. In \S~\ref{sc:simulation}, 
we perform simulations of collisional evolution of bodies from $r =
1$\,km. We show the dependence of embryo growth on turbulent strength
and collisional property and find the conditions for the formation of
cores massive enough for the onset of rapid gas
accretion to form gas giants. In \S~\ref{sc:discussion}, we discuss 
the growth timescale of embryos in turbulent disks, and 
the type I migration of growing embryos and the gas dispersal timescale
of disks. In \S~\ref{sc:sum}, we summarize our findings. 

\section{Critical Core Mass}
\label{sc:ccm}


The runaway growth of planetesimals 
produces a planetary embryo with mass $M_{\rm E}$ and
radius $R_{\rm E}$ in each annulus of a protoplanetary disk. 
Once $R_{\rm E}$ is larger than its Bondi radius,
$R_{\rm B}$
defined by $R_{\rm B} = G M_{\rm E}/c_{\rm s}^2$, the embryo has an
atmosphere, where $G$ is the
gravitational constant and $c_{\rm s}$ is the sound velocity of gas. 
The density of a hydrostatic atmosphere, $\rho_{\rm a}$, at the distance
$R$ from the center of the planetary embryo is given by
\citep[derivation in Appendix \ref{ap:atom_dens};][]{mizuno80,stevenson82,inaba_ikoma03}
\begin{equation}
 \rho_{\rm a} = \frac{\pi \sigma_{\rm SB} }{12 \kappa L_{\rm e} R^3} 
  \left(\frac{GM_{\rm E} \mu m_{\rm H}}{k}\right)^4,\label{eq:rho_a}
\end{equation}
where $\sigma_{\rm SB}$ is the Stefan-Boltzmann constant, $\kappa$ is
the opacity of the atmosphere, $k$ is the Boltzmann constant, $\mu$ is
the mean molecular weight, $m_{\rm H}$ is the mass of a hydrogen
atom, and $L_{\rm e}$ is the planetary luminosity. The accretion of
bodies onto planetary embryos mainly determines $L_{\rm e}$ so that 
\begin{equation}
 L_{\rm e} = \frac{GM_{\rm E}}{R_{\rm E}} \frac{dM_{\rm E}}{dt}. 
\end{equation}

The total atmospheric mass, $M_{\rm A}$, is given by 
\begin{eqnarray}
 M_{\rm A} &=& \int_{R_{\rm E}}^{R_{\rm B}} 4 \pi R^2 \rho_{\rm a} dR,\\
  &=& \frac{\pi^2 \sigma_{\rm SB} G^3 M_{\rm E}^3 R_{\rm E}}{3 \kappa
   \dot M_{\rm E} } 
  \left(\frac{\mu m_{\rm H}}{k}\right)^4 \ln\left(
\frac{R_{\rm B}}{R_{\rm E}} \right). \label{eq:matm}
\end{eqnarray}
Once $M_{\rm A}$ is comparable to $M_{\rm E}$, the hydrostatic
atmosphere is not maintained and then rapid gas accretion occurs to form
gas giants. The embryo mass at the onset of rapid gas accretion is
called the critical core mass, which is approximately estimated from 
$M_{\rm A}/M_{\rm E} \approx 1/3$ \citep{mizuno80,stevenson82,ikoma00}. 
Using Eq.~(\ref{eq:matm}) with $M_{\rm A} = M_{\rm E}/3$, the critical core mass $M_{\rm crit}$ is given by 
\begin{equation}
 M_{\rm crit} = 5 M_\oplus \left(\frac{ M_{\rm E} / \dot M_{\rm E}}{10\,{\rm
		     Myr}}\right)^{-3/4} 
 \left(\frac{\kappa}{0.01\,{\rm cm}^2 \,{\rm g}^{-1}}\right)^{3/4}, 
\end{equation}
where the dependence of $\ln (R_{\rm B}/R_{\rm E})$ on $M_{\rm E}$ is
ignored for this derivation. 

Therefore, forming embryos with $M_{\rm E} \ga 5M_\oplus$ are required for giant planet
formation via core accretion. 
Note that the critical core mass may be
smaller than $5M_\oplus$ because the depletion of planetesimals due to collisional
fragmentation results in smaller $\dot M_{\rm E}$
\citep[e.g.,][]{kobayashi11}. 

We below perform simulations
for the formation and growth of embryos and find conditions for the
formation of embryos with the critical core mass. 
Although we here simply estimate $M_{\rm A}$ using the
radiative temperature gradient with constant $\kappa$, 
we calculate $M_{\rm A}$ from $M_{\rm E}$ and $\dot M_{\rm E}$ obtained
in simulations, taking into account the convective temperature gradient
as well as the radiative one with $\kappa$ dependent on temperature \citep{inaba_ikoma03}. 

\section{Model for Planetary accretion}
\label{Sc:model}

The surface number density of
planetesimals with masses $m$ to $m+dm$ orbiting around the host star
with mass $M_*$ at the distance $a$, $n_{\rm s}(m,a) dm$, evolves through
collisions and radial drift. The governing equation is given as 
\citep[e.g.,][]{kobayashi16}, 
\begin{eqnarray}
 \frac{\partial m n_{\rm s}(m,a)}{\partial t} &=& \frac{m}{2} \int_0^\infty
  dm_1 \int_0^\infty dm_2 n_{\rm s}(m_1,a) n_{\rm s}(m_2,a) 
\nonumber\label{eq:coag}
\\
  && \quad \times K(m_1,m_2) \delta(m-m_1-m_2+m_{\rm e})
\nonumber
\\
&& - m n_{\rm s}(m) \int_0^\infty dm_2 n_{\rm s} (m_2,a) K(m,m_2) 
\nonumber
\\
&& + \frac{\partial}{\partial m} \int_0^\infty dm_1 \int_0^{m_1 } dm_2 
n_{\rm s}(m_1,a) n_{\rm s}(m_2,a) 
\nonumber
\\
 && \quad \times K(m_1,m_2) \Psi(m,m_1,m_2)
\nonumber
\\
&& - \frac{1}{a} \frac{\partial}{\partial a} [ a m n_{\rm s}(m,a) v_{\rm
 drift}(m,a)], 
\end{eqnarray}
where the collisional kernel $K(m_1,m_2)$ between bodies with masses $m_1$
and $m_2$ is given by 
\begin{equation}
 K(m_1,m_2) = 
(h_{m_1,m_2} a)^2 \langle {\cal P}_{\rm col}(m_1,m_2) \rangle \Omega, 
\end{equation}
with the reduced mutual Hill radius $h_{m_1,m_2} =
[(m_1+m_2)/3M_*]^{1/3}$, the dimensionless mean collisional rate
$\langle {\cal P}_{\rm col}(m_1,m_2) \rangle$, and the Kepler orbital frequency
$\Omega$, $m_{\rm e}$ and
$\Psi(m,m_1,m_2)$ are, respectively, the total and cumulative masses of
bodies produced by a single collision between bodies with masses $m_1$
and $m_2$, and $v_{\rm drift}$ is the drift velocity of a body due to
gas drag. The collisional rate 
is determined according to the orbits of colliding bodies and
the amount of their atmospheres: $\langle {\cal P}_{\rm col}(m_1,m_2) \rangle$ 
is given by a function of relative eccentricities and inclinations
between bodies and the accretion rates of the bodies, which are
summarized in \citet{inaba01}. The enhancement of collisional
cross-section due to embryo's atmosphere \citep{inaba_ikoma03} and due
to gas drag for small bodies \citep{ormel10b} is taken into account. 
The atmospheric opacity is given by the sum of
those of gas and grains, given by \citep{inaba_ikoma03}
\begin{equation}
 \kappa = \left\{
  \begin{array}{lll} 
   \displaystyle 
    0.01 + 4 f \,{\rm cm}^2\,{\rm g}^{-1} & {\rm for} & T \le 170\,{\rm K}, 
    \\
   \displaystyle
    0.01 + 2 f \,{\rm cm}^2\,{\rm g}^{-1} & {\rm for} &  170\,{\rm K} < T \le 1700\,{\rm K}, 
    \\
   \displaystyle
    0.01\,{\rm cm}^2\,{\rm g}^{-1} & {\rm for} &  T > 1700\,{\rm K},  
\end{array}
  \right.\label{eq:kappa}
\end{equation}
where $f$ is the grain depletion factor comparing to the interstellar
one. The effective growth of dust aggregates deplete the opacity
significantly \citep{okuzumi12}, so that we put $f=10^{-4}$. 

The collisional kernel depends on orbital eccentricities $e$ and
inclinations $i$ of colliders, which depend on mass $m$ and distance
$a$. 
The time variation of $e(m,a)$ and $i(m,a)$ via gravitational interactions
between bodies depends on the mass distribution of bodies. 
Therefore, the evolution of $e(m,a)$ and $i(m,a)$ 
have to be treated simultaneously with collisional evolution. 
The square of dispersions for eccentricities and inclinations of bodies
evolve according to 
\begin{eqnarray}
 \frac{d e^2(m,a)}{dt} &=& \left. \frac{d e^2}{dt}\right|_{\rm g}
  + \left. \frac{d e^2}{dt}\right|_{\rm d}
  + \left. \frac{d e^2}{dt}\right|_{\rm c}
  + \left. \frac{d e^2}{dt}\right|_{\rm t},\label{eq:de} \\
 \frac{d i^2(m,a)}{dt} &=& \left. \frac{d i^2}{dt}\right|_{\rm g}
  + \left. \frac{d i^2}{dt}\right|_{\rm d}
  + \left. \frac{d i^2}{dt}\right|_{\rm c}
  + \left. \frac{d i^2}{dt}\right|_{\rm t},\label{eq:di}
\end{eqnarray}
where the subscripts ``g'', ``d'', ``c'', and ``t'' indicate 
gravitational interaction between bodies \citep{ohtsuki02}, damping
by gas drag \citep{adachi76}, collisional damping
\citep{ohtsuki92}, and turbulent stirring due to density fluctuation
\citep{okuzumi13} and aerodynamical friction \citep{voelk80},
respectively. 
We use the damping rates by gas drag linear
functions of $e$ and $i$ given by \citet{inaba01} because $e$ and $i \ll
1$, although the $e$ and $i$ damping rates by gas drag are greater 
for higher $e$ and $i$ \citep{kobayashi15}.
We use the damping rate and radial drift due to gas drag, taking into
account three gas drag regimes \citep[see details in][]{kobayashi+10}. 

For 10 meter-sized or larger bodies, the stirring by turbulence is
mainly caused by turbulent density fluctuation, given by
\citep{okuzumi13} 
\begin{eqnarray}
 \left. \frac{d e^2}{dt}\right|_{\rm t} &\approx& f_{\rm d} 
  \left(
   \frac{\Sigma_{\rm g} a^2}{M_*}
  \right)^2
  \Omega, \\
\left. \frac{d i^2}{dt}\right|_{\rm t}  &\approx& \epsilon^2
 \left. \frac{d e^2}{dt}\right|_{\rm t}, 
\end{eqnarray}
where $\Sigma_{\rm g}$ is the gas surface density, $f_{\rm d}$ is the
dimensionless factor, and $\epsilon = 0.01$ \citep[see ][]{kobayashi16}. According to the magnetohydrodynamical
simulations, $f_{\rm d}$ is given by \citep{okuzumi13}, 
\begin{equation}
 f_{\rm d} = \frac{0.94 \alpha}{(1+4.5 H_{\rm res,0}/H)^2},\label{eq:fd}
\end{equation}
where $H$ is the scale height of the disk, $H_{\rm res,0}$ is the half
vertical width of the dead zone, $\alpha$ is the dimensionless turbulent
viscosity at the midplane. For simplicity, we set $H_{\rm res,0} = H$. 

\begin{figure}[htbp]
 \epsscale{1.} \plotone{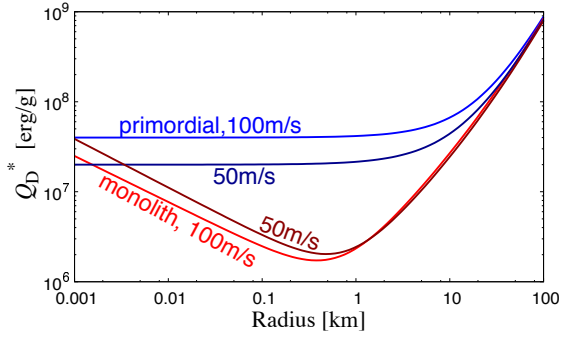} 
 \figcaption{The fragmentation energy $Q_{\rm D}^*$ for
 primordial and monolith bodies with impact velocities of 50\,m/s and 100\,m/s, given in
 Eqs.~(\ref{eq:qds})--(\ref{eq:qs_porous}). 
\label{fig:qds}
} 
\end{figure}

Dynamical stirring by large bodies increases collisional
velocities of bodies. The collisional outcome is determined by $m_{\rm
e}$ and $\Psi(m,m_1,m_2)$ in Eq.~(\ref{eq:coag}), which are mainly
controlled by the fragmentation energy $Q_{\rm D}^*$ (see detailed setting Appendix~\ref{app_collisional_modeling}). 
We use the following formula for $Q_{\rm D}^*$; 
\begin{equation}
  Q_{\rm D}^* = Q_{\rm 0s} \Biggl(\frac{r}{1\,{\rm cm}}\Biggr)^{\beta_{\rm s}} +
Q_{\rm 0g} \rho_{\rm s} \Biggl(\frac{r}{1\,{\rm cm}}\Biggr)^{\beta_{\rm g}} +
C_{\rm gg}
\frac{2Gm}{r}, \label{eq:qds} 
\end{equation}
where 
$\rho_{\rm s}$ is the density of a body, $Q_{\rm 0s}, Q_{\rm 0g},\beta_{\rm s}$, $\beta_{\rm
g}$, and $C_{\rm gg}$ characterize the collisional strength. 
On the right hand side of Eq.~(\ref{eq:qds}), the
first term is dominant for
kilometer sized or smaller bodies. The second term is important for larger
bodies. The third term is determined by pure gravity, which is dominant
for $r \ga 10^7$\,cm: we set $C_{\rm gg} = 10$ \citep{stewart09}. 
The values of $Q_{\rm 0s}, Q_{\rm 0g},\beta_{\rm s}$, and $\beta_{\rm
g}$ depend on the impact velocity \citep[e.g.,][]{benz99} and the
structure of a body \citep[e.g.,][]{wada13}. 
For monolith bodies, these values are estimated from the interpolation
of simulation data for the impact velocities $v_{\rm imp}$ of 3 and $5\,{\rm km/s}$
obtained by \citet{benz99},
given by 
\begin{eqnarray}
 Q_{\rm 0s} &=& 1.6 \times 10^7
 \left( \frac{v_{\rm imp}}{3\,{\rm km/s}} \right)^{\gamma_{\rm sv}} \,{\rm
 erg/g},\label{eq_qss} 
\\
\beta_{\rm s} &=& -0.39 
\left(\frac{v_{\rm imp}}{3\,{\rm km/s}}\right)^{\gamma_{\rm bsv}},\label{eq:qs_betas}\\
 Q_{\rm 0g} &=& 1.2 
\left(\frac{v_{\rm imp}}{3\,{\rm km/s}}\right)^{\gamma_{\rm gv}} \,{\rm
erg \, cm}^3 {\rm /g}^{2},\label{eq:q0g} \\
\beta_{\rm g} &=& 1.26 
\left(\frac{v_{\rm imp}}{3\,{\rm km/s}}\right)^{\gamma_{\rm bgv}},\label{eq:qs_betag}
\end{eqnarray}
where $\gamma_{\rm sv} = -0.82$, $\gamma_{\rm gv} = -0.31$, $\gamma_{\rm
bsv} = -0.080$, and $\gamma_{\rm bgv} = 0.032$. 
However, growing bodies via collisions are porous
\citep[e.g.,][]{wada13}, which might not have monolith-like
structures until melting. Therefore, bodies smaller than about 10\,km
have $Q_{\rm D}^*$ of porous bodies, which is given by \citep{wada13},
\begin{eqnarray}
 Q_{\rm s0} &=& 4 \times 10^7 
 \left( \frac{v_{\rm imp}}{100 \,{\rm m/s}} \right) \,{\rm
 erg/g}, \label{eq:qs_porous} \\
 \beta_{\rm s} &=& 0. 
\label{eq:betas_porous} 
\end{eqnarray}
We investigate the growth of melted bodies (monolith) and primordial
bodies. 
For melted bodies, $Q_{\rm D}^*$ is given by
Eqs.~(\ref{eq_qss})--(\ref{eq:qs_betag}). 
For primordial bodies, $Q_{\rm D}^*$ is calculated from 
Eqs. (\ref{eq:q0g})--(\ref{eq:betas_porous}). 
For melted and primordial bodies, 
$Q_{\rm D}^*$ are shown in Fig. \ref{fig:qds}. 

\section{Mass Evolution of Bodies}
\label{sc:simulation}

We perform simulations for planet formation in disks from 4.8 to
26\,AU via the time integration of
Eqs.~(\ref{eq:coag}), (\ref{eq:de}), and (\ref{eq:di}) for $M_* =
M_\sun$, where $M_\sun$ is the solar mass. 
The disk is divided into 10 annuli and the
mass distribution is described using mass bins with radios 
between adjacent mass bins 1.05. 
We fix the bulk density of bodies at $\rho_{\rm s} = 1 \,{\rm g/cm}^3$; the mass-radius
relation is given by $m = 4 \pi \rho_{\rm s} r^3/3$. 
The mass corresponding to the smallest mass bin is set to $4.2\times
10^6$\,g ($r = 1$\,m). 
We set the initial bodies at radius
of 1\,km. The bodies initially have $e = i /\sqrt{\epsilon} = 3
v_{\rm esc}/a \Omega \approx 9.9 \times 10^{-5}$. 
\revhk{The collisional growth of bodies is almost independent of the
initial radius and orbits of bodies if the new born planetesimals are
smaller than those at the onset of runaway growth \citep{kobayashi16}.
}

\begin{figure}[thbp]
 \epsscale{1.} \plotone{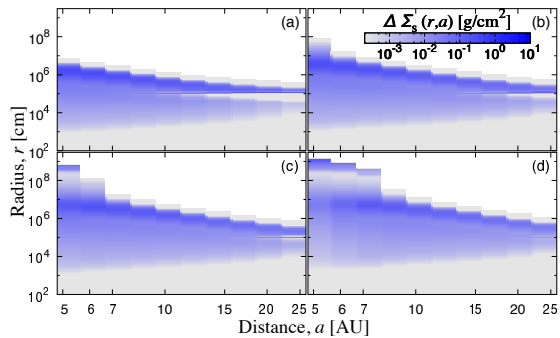} \figcaption{Size
 distribution for mass surface density $\Delta \Sigma_{\rm s}(r,a)$ of bodies 
at $2.2 \times
 10^5$ yr (a), $7.2 \times 10^5$ yr (b), $1.4\times 10^6$ yr
 (c), and $3.3\times 10^6$ yr (d) 
in a disk with $x_{\rm s} = x_{\rm
 g} = 3$ for $\alpha = 3 \times 10^{-3}$ with collisional fragmentation for
 primordial collisional strength, as a function of the distance from the
 host star, $a$, and the radius of bodies corresponding to their mass $m$. 
\label{fig:mass_cont} }
\end{figure}

\begin{figure}[htbp]
 \epsscale{1.} \plotone{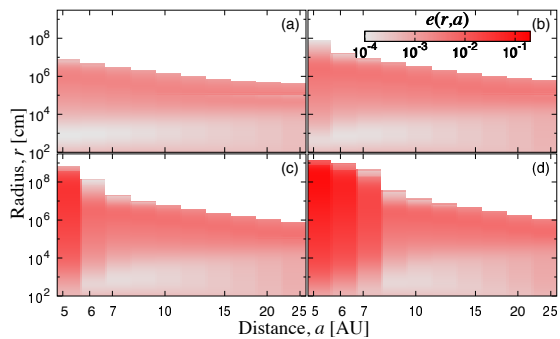} \figcaption{
Orbital eccentricity distribution obtained from the same simulation as
 Fig.~\ref{fig:mass_cont}. 
 \label{fig:ecc_cont} }
\end{figure}

\begin{figure}[htbp]
 \epsscale{1.} \plotone{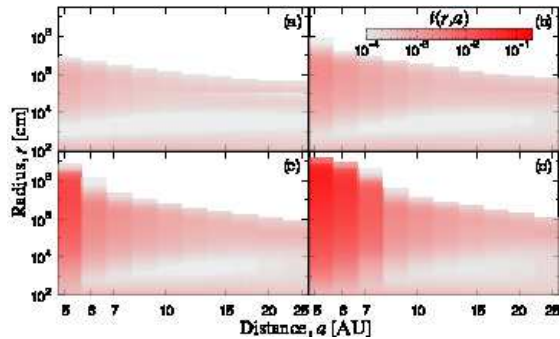} 
\figcaption{
Orbital inclination distribution obtained from the same simulation as
 Fig.~\ref{fig:mass_cont}. 
\label{fig:inc_cont} }
\end{figure}

The initial surface densities of solid and gas, $\Sigma_{\rm g,0}$ and
$\Sigma_{\rm s,0}$, are, respectively, set to have power-law radial
distributions; 
\begin{eqnarray}
\Sigma_{\rm g,0} &=& 1700 x_{\rm g}\left(\frac{a}{\rm
				    1\,AU}\right)^{-1.5} \,{\rm g\,
cm}^{-2},
\\
 \Sigma_{\rm s,0} &=& 30 x_{\rm s} \left(\frac{a}{\rm
				    1\,AU}\right)^{-1.5} \,{\rm g\,
 cm}^{-2},
\end{eqnarray}
where $x_{\rm g}$ and $x_{\rm s}$ are the scaling factors and the disk
with $x_{\rm g} = x_{\rm s} = 1$ corresponds to the minimum-mass solar
nebula (MMSN) model \citep{hayashi81}. 
The solid surface density $\Sigma_{\rm s}$ 
evolves due to the radial drift of bodies, while the gas surface density
artificially decreases from the formula $\Sigma_{\rm g} = \Sigma_{\rm g,0}
\exp(-t/\tau_{\rm gas})$ with $\tau_{\rm gas} = 10^7\,{\rm yr}$. 
We discuss $\tau_{\rm gas}$ and the time dependence of depletion in
\S~\ref{sc:discussion}. 
\revhk{In addition, we set the temperature at the disk midplane as 
\begin{equation}
 T = 280 \left( \frac{a}{1\,{\rm AU}}\right){\rm K}, 
\end{equation}
which affects $R_{\rm B}$ and the aerodynamical turbulent stirring. 
However, they are unimportant for the evolution of $M_{\rm E}$ and
$M_{\rm A}$. 
}

We show the size and velocity evolution of bodies in the whole disk
in \S~\ref{sc:color_dist} and focus on the size and velocity
distributions around 5\,AU in \S~\ref{sc:dist_5au}. The evolution of
planetary embryos dependent on $\alpha$ is shown in
\S~\ref{sc:mlarge_alpha}.  
We discuss the formation of cores with the critical core masses from $M_{\rm E}$ and $M_{\rm
A}$ obtained within disk lifetimes by simulations in
\S~\ref{sc:matm}. 

\subsection{Size and Velocity Evolution of Bodies in the Whole Disk}
\label{sc:color_dist}

We perform a simulation for planet formation with primordial
strength in the disk with $x_{\rm s} = x_{\rm g} = 3$ (3MMSN) for
$\alpha = 3 \times 10^{-3}$, which results in the size and orbit
evolution of bodies as shown in
Figs.~\ref{fig:mass_cont}-\ref{fig:inc_cont}.  The surface density of
solid bodies is given by $\Sigma_{\rm s}(a) = \int m n_{\rm s}(m,a) dm = \int
m^2 n_{\rm s}(m,a) d \ln m$, so that we define 
the surface density of bodies in the logarithmic mass interval as
\begin{equation}
 \Delta \Sigma_{\rm s} (r,a) \equiv m^2 n_{\rm s}(m,a), 
\end{equation} 
with $r$ corresponding to $m$. 
Fig.~\ref{fig:mass_cont} shows $\Delta \Sigma_{\rm s}(r,a)$. 
The collisional growth of bodies occurs
inside-out. Orderly growth initially occurs so that $\Delta \Sigma_{\rm s}$ have
a maximum around radii of largest bodies in each annulus for $t \la
1$\,Myr (Fig.~\ref{fig:mass_cont}a,b).  Larger bodies tend to have greater $e$
and $i$ for $r \ga 10$\,m (Fig.\ref{fig:ecc_cont}a,b and
\ref{fig:inc_cont}a,b).  Density-fluctuation turbulent stirring and gas
damping mainly control $e$ and $i$ for $r \ga 10$\,m, while the
fluid-dynamical turbulent stirring is dominant instead of that by
density fluctuation for $r \la 10$\,m.  The runaway growth of bodies
produces planetary embryos inside 8\,AU by 4\,Myr
(Fig.~\ref{fig:mass_cont}c,d). The stirring by embryos increases
$e$ and $i$ (Figs.~\ref{fig:ecc_cont}c,d and
\ref{fig:inc_cont}c,d), which induces collisional fragmentation due to
high speed collisions. Collisional fragments become smaller due to further
collisions between themselves until radial drift, which 
 reduces the surface
density of bodies with $r \la 10$ -- 100\,km (compare
Figs.~\ref{fig:mass_cont}c and \ref{fig:mass_cont}d). 
Planetary embryo formation in the outer disk induces 
collisional fragmentation and radial drift, which supplies small bodies
in the inner disk. However, the net flux of small bodies reduces the
surface density of bodies around planetary embryos in the inner disk. 
This stalls the
growth of embryos \citep[e.g.,][]{kobayashi+10,kobayashi11}.  
In order
to see detailed mass and orbit evolution, we focus on the evolution at
5.2\,AU in the following paragraphs.

\begin{figure}[thbp]
 \epsscale{1.} \plotone{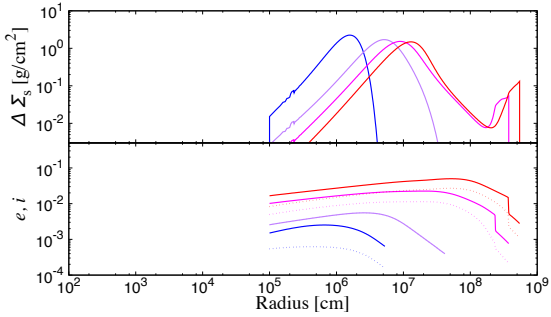} \figcaption{Size
 distribution of $\Delta \Sigma_{\rm s}$ of bodies (top) and their eccentricities $e$ (bottom;
 solid curves) and
 inclinations $i$ (bottom; dotted curves) at 5.2\,AU in a disk with
 $x_{\rm s} = x_{\rm g} = 3$ for $\alpha = 3 \times 10^{-3}$ without collisional fragmentation. The curves indicate those at $1.7 \times
 10^5$ yr (blue), $8.7 \times 10^5$ yr (purple), $1.9\times 10^6$ yr
 (magenta), and $3.2\times 10^6$ yr (red). The time evolution results in
 the existence of large bodies, so that the curves move from left to
 right.  \label{fig:mass_evo_nofrag} }
\end{figure}
\begin{figure}[htbp]
 \epsscale{1.} \plotone{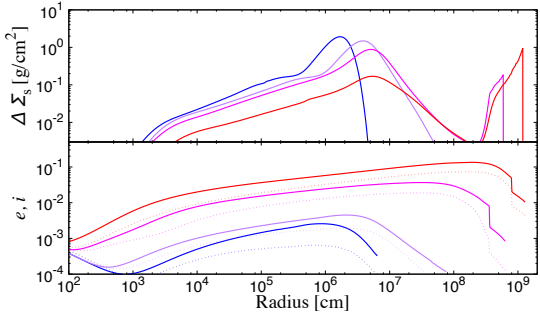} \figcaption{
Same as Fig.\ref{fig:mass_evo_nofrag} but with collisional fragmentation
 for primordial strength. The curves indicate those at $2.2 \times
 10^5$ yr (blue), $7.2 \times 10^5$ yr (purple), $1.4\times 10^6$ yr
 (magenta), and $3.3\times 10^6$ yr (red). 
 \label{fig:mass_evo_Qp} }
\end{figure}
\begin{figure}[htbp]
 \epsscale{1.} \plotone{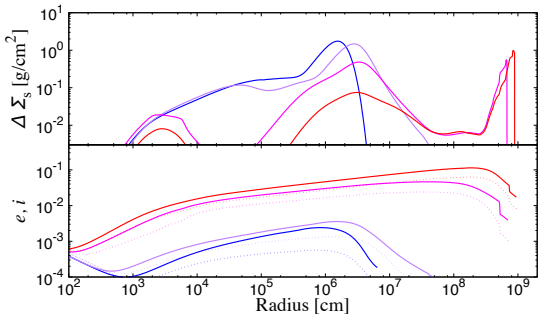} \figcaption{
Same as Fig.\ref{fig:mass_evo_nofrag} but with collisional fragmentation
 for monolith strength. The curves indicate those at $2.1 \times
 10^5$ yr (blue), $4.9 \times 10^5$ yr (purple), $1.1\times 10^6$ yr
 (magenta), and $3.1\times 10^6$ yr (red). 
 \label{fig:mass_evo_Qm} }
\end{figure}

\subsection{Size and Velocity Distributions around 5\,AU}
\label{sc:dist_5au}

As shown in Figs.~\ref{fig:mass_cont}--\ref{fig:inc_cont}, 
collisional fragmentation naturally occurs during planet formation and 
plays an important role for planet formation
\citep[e.g.,][]{wetherill93,inaba03,kobayashi+10,kobayashi11,kobayashi12,kobayashi13}. Therefore collisional fragmentation is
necessary to be taken into account for planet formation. 
For comparison, however, we first show the size and orbit evolution of bodies 
in the case without collisional fragmentation ($m_{\rm e} = \Psi =0$ in
Eq. \ref{eq:coag}).  
Fig.~\ref{fig:mass_evo_nofrag} shows the evolution of $\Delta
\Sigma_{\rm s}$,
$e$, and $i$ of bodies at 5.2\,AU in the 3MMSN disk with $\alpha = 3
\times 10^{-3}$, as a function of $r$ corresponding to $m$. 
For $t \la 1$\,Myr, the surface densities $\Delta \Sigma_{\rm s}(r)$
have single peaks, which move to large $r$. 
This is caused
by orderly growth of bodies. 
The peak radius $r_{\rm pk}$
is approximated to be the mass-weighted average radius of bodies. 
For $t \ga 1$\,Myr, the size distribution becomes wider and then
collisional evolution 
produces another peak at the high-mass end, which indicates planetary
embryos. This is caused by the runaway growth of bodies. After the onset
of runaway growth, $r_{\rm pk}$ does not change significantly so that 
the peak radius at the onset of runaway growth, $r_{\rm rg}$, is
approximated to be $r_{\rm pk}$, which 
is
estimated to $\sim 100\,$km from the size distribution of bodies at
$\sim 1$\,Myr. 
The onset of
runaway growth happens if $v_{\rm r} \la 1.5 v_{\rm esc}$ for bodies of
$r \sim r_{\rm pk}$ \citep{kobayashi16}. 
Since $v_{\rm r} / a \Omega \approx e$ and $v_{\rm esc}/a \Omega
\approx 6\times 10^{-3} (r/100\,{\rm km})$, we estimate 
$r_{\rm rg} \approx 100{\rm \,km}$ from the data of $e$ at 1.9\,Myr in
Fig.~\ref{fig:mass_evo_nofrag}, which is consistent with the mass
distribution in Fig.~\ref{fig:mass_evo_nofrag}. 

The size distributions of bodies have single peaks at $r = r_{\rm pk}$ 
prior to runaway growth, $t \la 1$\,Myr. 
Bodies with $r \la r_{\rm pk}$ have similar slopes of the size
distributions of bodies. 
Even after runaway growth, bodies with $r \la r_{\rm rg}$ have similar
slopes. The slope of the mass distribution of bodies is estimated by eye
to $d \ln \Delta \Sigma_{\rm s} (r) / d \ln r \approx 1.8$. 
For collisional cascade with $Q_{\rm D}^*$ and $v_{\rm r}$ independent
of $m$, $d \ln n_{\rm s}(m) / d \ln m = 1/2$ \citep{dohnanyi,tanaka96}. 
The slopes obtained from simulations without collisional fragmentation
are steeper than that of the simple collisional cascade. On the other
hand, the slopes of bodies with $r$ ranging from $\approx r_{\rm rg}$ to
$10^5$\,km are $d \ln \Delta \Sigma_{\rm s}(r) / d \ln r \approx -2.0$
in $t \ga 1$\,Myr. 
The collisional cross-section is proportional to $m^{5/3}/v_{\rm r}^2$ due to
gravitational focusing and 
$v_{\rm r} \propto m^{-1/2}$ due to dynamical friction, which
analytically gives $d \ln \Delta \Sigma_{\rm s} (r) / d \ln r = -2$
\citep{makino}. 
The slopes formed via runaway growth is roughly explained by
gravitational focusing and dynamical friction. In more detailed
analysis, the runaway-growth slopes bend slightly \citep[see][]{morishima17}. 

Fig. \ref{fig:mass_evo_Qp} shows the evolution of bodies at 5.2\,AU in the same
condition as Figs.~\ref{fig:mass_cont}-\ref{fig:inc_cont} (i.e., the
case with collisional fragmentation for primordial strength). 
For $t \la 1$\,Myr, 
$\Delta \Sigma_{\rm s} (r)$ have a single peak, although $\Delta \Sigma_{\rm s} (r)$ have a long tail at
low-mass side, which is produced by collisional fragmentation. 
These small bodies make collisional damping effective, which reduces
$v_{\rm r}$. 
Runaway growth occurs slightly earlier than the case
without collisional fragmentation, resulting in $r_{\rm rg} \approx
60$\,km smaller than $r_{\rm rg}$ without fragmentation. 
After the onset of runaway growth ($t \ga 1$\,Myr), $r_{\rm pk}$ moves insignificantly. 
The stirring by planetary embryos created by runway growth increases
$e$ and $i$ of surrounding planetesimals, which induces collisional
fragmentation of planetesimals of $r \sim r_{\rm rg}$. The solid surface density
 is decreased via collisional fragmentation and radial
drift of resultant fragments \citep{kobayashi+10,kobayashi11}. Planetary
embryos grow through the accretion of planetesimals and fragments until
their depletion. 

The size distributions of bodies have single peaks at $r = r_{\rm pk}$ prior to runaway
growth, which is similar to the case without collisional fragmentation. 
However, bodies smaller than the peaks have multiple slopes:  
The slopes are estimated by eye to $d \ln \Delta \Sigma_{\rm s} (r) / d \ln r \approx 2.5$ for $r
\ll 30$\,m, $d \ln \Delta \Sigma_{\rm s} (r) / d \ln r \approx 0.6$ from $r \approx
30$\,m to $\sim 0.2 r_{\rm pk}$, and $d \ln \Delta \Sigma_{\rm s} (r) / d \ln r
\approx 2.0$ for $r \approx 0.2 r_{\rm pk}$\,--\,$r_{\rm pk}$ (see Fig.~\ref{fig:mass_evo_Qp}). 
The slope for large bodies of $r = 0.2 r_{\rm pk}$ -- $r_{\rm pk}$ is similar to that for no fragmentation,
which is simply determined by collisional growth of bodies in orderly
growth. 
The intermediate sized bodies for $r = 30$\,m -- $0.2 r_{\rm pk}$, which are mainly produced by erosive
collisions of bodies with $r \sim r_{\rm pk}$,  
have so small $e$ and $i$ that
collisional fragmentation is negligible. The collisional growth among
them results in the slope determined by collisional cascade to the
positive mass direction, $d \ln \Delta \Sigma_{\rm s}(r) / d \ln r = 1/2$ \citep{tanaka96}. 
For bodies with $r \ll 30$\,m, radial drift as well as
the collisional cascade affects the mass distribution so that the slope
is given by $d \ln \Delta \Sigma_{\rm s}(r) / d \ln r =
5/2$\footnote{\revhk{Such bodies feel gas drag in the Stokes regime so
that $v_{\rm r} \propto r^{-2}$. The collisional cascade gives $(d \ln
\Delta \Sigma_{\rm s}(r) / d \ln r)_{\rm cas} = 1/2$ and the modulation due to
radial drift results in $d \ln
\Delta \Sigma_{\rm s}(r) / d \ln r = (d \ln \Delta \Sigma_{\rm s}(r) / d \ln
r)_{\rm cas} - d \ln v_{\rm r} / d \ln r $. }}. 
On the other hand, the runaway growth produces a
different power-law size distribution of bodies with $r\ga r_{\rm rg}$;
$d \ln \Delta \Sigma_{\rm s}(r)/d \ln r = -2.0$, which is caused by runaway growth similar
to the case without fragmentation. For $r \la r_{\rm rg}$, the slope is controlled by collisional
fragmentation due to large $e$ and $i$. 
Collisional cascade therefore occurs,
resulting in 
$d \ln \Delta \Sigma_{\rm s} (r) / d \ln r = (1+3p) / (2+p)$ with $3p = d \ln
Q_{\rm D}^* / d \ln r  - 2 d \ln v_{\rm r} / d \ln r$ 
\citep{kobayashi10}. 
Although $d \ln Q_{\rm D}^* / d \ln r$ depends on collisional velocities, $p \approx 0$ is roughly
estimated and then $d \ln \Delta \Sigma_{\rm s} (r) / d \ln r \approx
0.5$, which is similar to the slope for $r \sim 30$\,m -- 50\,km. 
For $r\la 30$\,m, bodies have so small $e$ and
$i$ that collisional fragmentation no longer occurs. The slope is
determined by collisional growth and radial drift, which is almost the same as the
slope of small bodies prior to runaway growth. 

Fig.~\ref{fig:mass_evo_Qm} shows the result of another simulation with monolith strength for the same disk
condition. The evolution of size distribution is
similar to the case for primordial strength; but the onset of
runaway growth occurs slightly earlier, resulting in $r_{\rm rg} \approx
30$\,km. This is because the small bodies produced by collisional
fragmentation prior to runaway growth results in slightly effective
collisional damping. After the onset of runaway growth ($t \ga 1$\,Myr), 
forming planetary embryos induce collisional fragmentation of
planetesimals, which mainly produces bodies with $r \sim 10$\,m--10\,km
through collisional cascade. 
The surface density $\Delta \Sigma_{\rm s} (r)$ has a
peak at the radius of $10$--100\,m. 
This is because the collisional cascade starting from collisional
fragmentation of bodies with $r \sim r_{\rm rg}$ is stalled by $e$ and
$i$ damping by gas drag in the Stokes regime for bodies with $r \la  10$
-- 100\,m \citep{kobayashi+10}. The magnitudes of
such peaks depend on $r_{\rm rg}$, $Q_{\rm D}^*$, $\Sigma_{\rm s}$, and
so on. Planetary embryos mainly grow through collisions with bodies of $\sim
r_{\rm rg}$ or $10$ -- 100\,m \citep{kobayashi+10,kobayashi11}. 
On the other hand, $\sim 300$\,m-sized bodies are quickly destroyed via
collisions with small bodies because of low $Q_{\rm D}^*$. 

The mass distribution prior to runaway growth is similar to the case
with primordial strength. The slopes of small bodies are almost the same,
while the population of small bodies is larger because of a high
production caused by collisional fragmentation due to low $Q_{\rm
D}^*$. On the other hand, although runaway growth results in a slope for $r \ga
30\,{\rm km} \approx r_{\rm rg}$ similar to the case with primordial
strength, bodies with $r \la
r_{\rm rg}$ have a ``wavy'' structure, where the slopes $d \ln \Delta
\Sigma_{\rm s} (r)
/ d \ln r$ varies in small size ranges; $d \ln \Delta \Sigma_{\rm s}(r)
/ d \ln r \approx 0.1$ for $r \approx 3$--60\,km, $d \ln \Delta
\Sigma_{\rm s}(r) / d \ln r \approx 0.2$ for $r\approx 0.3$--3\,km, $d
\ln \Delta \Sigma_{\rm s}(r) / d \ln r \approx -0.2$ for $r \approx
30$--300\,m, and $d \ln \Delta \Sigma_{\rm s}(r) / d \ln
r \approx 0.2$ for $r \la 30$\,m ($t = 1.9$\,Myr in
Fig.~\ref{fig:mass_evo_Qm}). 
The slope controlled by
collisional cascade is analytically estimated to $d \ln \Delta \Sigma_{\rm s}(r) / d
\ln r \approx 0.1$ for $r \ga 500$\,m and $d \ln \Delta \Sigma_{\rm
s}(r) / d \ln r \approx 0.0$ for $r \approx 30$--500\,m. Collisional
cascade may explain the slope only around $r \approx 3$--60\,km. The ``wavy'' pattern
for $r \la 3$\,km is formed due to large $v_{\rm r}^2/Q_{\rm D}^*$
\citep[e.g.,][]{campobagatin94,durda,thebault03,krivov07,loehne08}. The
value of $v_{\rm r}^2/Q_{\rm D}^*$ becomes significant around $r =
500$\,m because of minimum $Q_{\rm D}^*$ (see Fig.~\ref{fig:qds}), which
produces a bump at $r \approx 500$\,m in the size distribution.

\begin{figure}[htbp]
 \epsscale{1.} \plotone{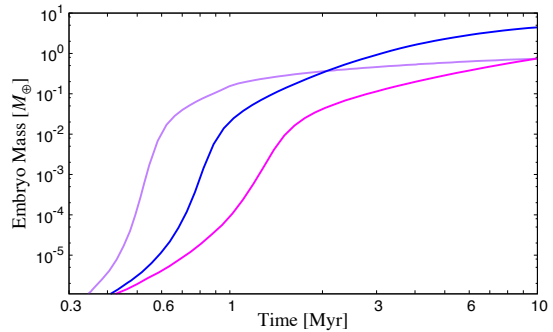} \figcaption{
Time evolution of planetary-embryo masses for no fragmentation
 (magenta), monolith strength (purple), and primordial strength (blue). 
  \label{fig:growth_g3a1e-4} }
\end{figure}

Fig.~\ref{fig:growth_g3a1e-4} shows growth of planetary
embryos obtained from simulations shown in
Figs. \ref{fig:mass_evo_nofrag}--\ref{fig:mass_evo_Qm}.  
For $t \la
0.4$ -- 1\,Myr, embryo masses $M_{\rm E}$ grow as $M_{\rm E} \propto t^3$
because of orderly growth \citep{kobayashi16}. The onset of runaway
growth occurs around 0.4 -- $1$\,Myr, resulting in strong time dependence of
$M_{\rm E}$.  The rapid growth occurs until $M_{\rm E} \sim
10^{-2}$--$10^{-1}\,M_\oplus$. The growth timescale in runaway growth
depends on the collisional model, which is caused by different $r_{\rm
rg}$.  After the rapid growth, slow growth occurs again, which is called
oligarchic growth. Planetary embryos grow through surrounding
planetesimals whose typical radii are $r_{\rm rg}$ (see
Figs.~\ref{fig:mass_evo_nofrag}--\ref{fig:mass_evo_Qm}). The viscous
stirring by embryos increases $e$ and $i$ of planetesimals, which
induces collisional cascade. Resultant bodies of $\sim 10$ -- 100\,m
drift inward rapidly. Therefore, the oligarchic growth is stalled by the
depletion of planetesimals due to collisional fragmentation and radial
drift of fragments. The growth of embryos at the oligarchic stage depends on
$Q_{\rm D}^*$ for $r \la r_{\rm rg}$ that controls the depletion of the
solid surface density (see
Figs.~\ref{fig:mass_evo_Qp} and \ref{fig:mass_evo_Qm}). 
The oligarchic growth for primordial strength 
stalls later than that for melted materials. 
For primordial strength, 
the characteristic radii of planetesimals, $r_{\rm rg}$, are large 
due to the late onset of runaway growth, which have great $Q_{\rm
D}^*$. The insignificant collisional depletion of planetesimals results
in the long accretion of planetesimals onto embryos so that  
embryos grow to be massive. In addition, collisional cascade grinds
planetesimals down to 10 -- 100 \,m. 
Strong gas drag damps $v_{\rm r}$ for $r \la 100$\,m so that collisional
cascade stalls without destructive collisions. 
The radius of bodies at the low mass end of collisional cascade, $r_{\rm
cc}$, 
depends on $Q_{\rm D}^*$; primordial strength has large $r_{\rm cc}$. 
Bodies with $r \sim r_{\rm cc}$ have so small $v_{\rm r}$ that they
effective accrete onto embryos, while their drift timescales are short
because of strong gas drag. The accretion efficiency of bodies with $r
\sim r_{\rm cc}$ depends on $r_{\rm cc}$ \citep{kobayashi+10}. For
primordial material, embryos effectively grow via accretion of bodies with $r \sim
r_{\rm cc}$ because of slow radial drift for large $r_{\rm cc}$ due to
great $Q_{\rm D}^*$. 

\subsection{Turbulent Strength Dependence of Embryo Growth}
\label{sc:mlarge_alpha}

\begin{figure}[htbp]
 \epsscale{1.} \plotone{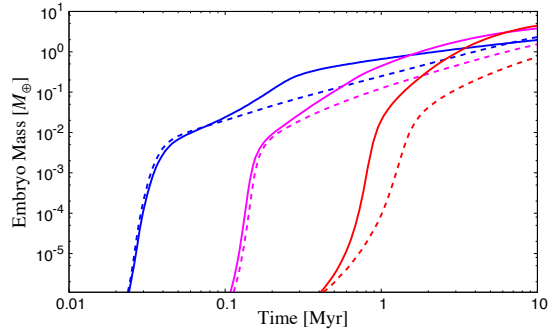} \figcaption{
Time evolution of planetary-embryo masses in 3MMSN disk with $\alpha = 3
 \times 10^{-5}$
 (blue), $3 \times 10^{-4}$ (magenta), and $3 \times 10^{-3}$ (red) for collisional model
 with primordial $Q_{\rm D}^*$ (solid) and without fragmentation
 (dotted). 
  \label{fig:growth_alpha_prim} }
\end{figure}

We carry out the collisional-evolution simulations in 3MMSN disks with
$\alpha = 3 \times 10^{-5}$ -- $3 \times 10^{-3}$ for primordial $Q_{\rm D}^*$, comparing
with the results without collisional fragmentation (see
Fig.~\ref{fig:growth_alpha_prim}). For weak turbulence ($\alpha \la 
10^{-4}$), the growth of embryos is similar to the case
without fragmentation until embryos are larger than the Moon mass ($\sim
10^{-2} M_\oplus$). 
For $\alpha = 3  \times 10^{-3}$, 
the early collision fragmentation due to strong turbulent stirring 
leads to effective collisional damping,
inducing the onset of runaway growth earlier comparing to the case
without collisional fragmentation. 
After runaway growth, 
large embryos controls $v_{\rm r}$ instead of turbulence, which 
 induces significant collisional
fragmentation of planetesimals. Collisional cascade results in bodies
of $r \sim r_{\rm cc} \sim 10$ --
100\,m. The effective accretion of such small bodies leads to the
rapid growth of embryos (compare the results with or without collisional
fragmentation). However, the growth is stalled by the depletion
of small bodies due to collisional fragmentation and radial drift. 

For small $\alpha$ the runaway growth occurs early, resulting in small
$r_{\rm rg}$. If we ignore collisional fragmentation, the early
formation and growth of
embryos due to small $\alpha$ results in large embryos (see
Fig.~\ref{fig:growth_alpha_prim}). However, collisional fragmentation
affects the growth of embryos.  Small planetesimals tend to be
destroyed via collision due to small $Q_{\rm D}^*$ for weak
self-gravity (see Fig.~\ref{fig:qds}).  The stirring by small embryos
increases $v_{\rm r}$ of planetesimals moderately, which can induces
collisional fragmentation of planetesimals if small $r_{\rm rg}$.  Resultant fragments accelerate the growth of embryos initially, while
the depletion of surrounding planetesimals 
stalls embryo growth.  Weak turbulence (small $\alpha$) results in
small masses of embryos in the late stage ($t \ga 4$\,Myr in
Fig.~\ref{fig:growth_alpha_prim}), while large $\alpha$ enhances $r_{\rm
rg}$ and then $M_{\rm E}$ at the late stage. 

\begin{figure}[htbp]
 \epsscale{1.0} \plotone{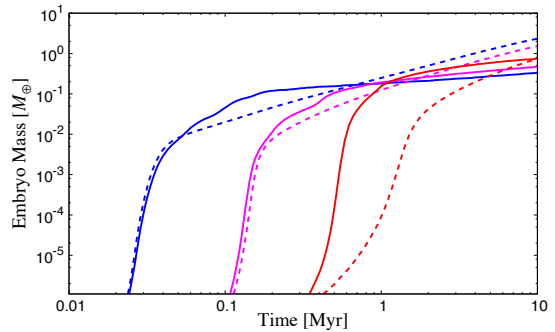} \figcaption{
Same as Fig.~\ref{fig:growth_alpha_prim}, but for monolith $Q_{\rm D}^*$
 (solid). 
  \label{fig:growth_alpha_mono} }
\end{figure}

Fig.~\ref{fig:growth_alpha_mono} shows embryo growth for monolith
$Q_{\rm D}^*$. The $\alpha$ dependence is similar to the result for
primordial $Q_{\rm D}^*$; higher $\alpha$ produces more massive embryos. 
However, final embryos are smaller. 
The accretion of small bodies with $r \sim r_{\rm cc} \approx 10$ --
100\,m contributes to embryo growth. However, the depletion of bodies with
$r \sim r_{\rm cc}$ due to radial drift stalls embryo growth. 
Low $Q_{\rm D}^*$ for $r \sim 10$ -- 100\,m makes $r_{\rm cc}$
small. Therefore, the growth for monolith strength stalls earlier 
than that for the primordial strength (see Figs.~\ref{fig:growth_alpha_prim} and
\ref{fig:growth_alpha_mono}). 

\subsection{Forming Cores with Critical Core Masses}
\label{sc:matm}

\begin{figure*}[htbp]
 \epsscale{.45} 
\plotone{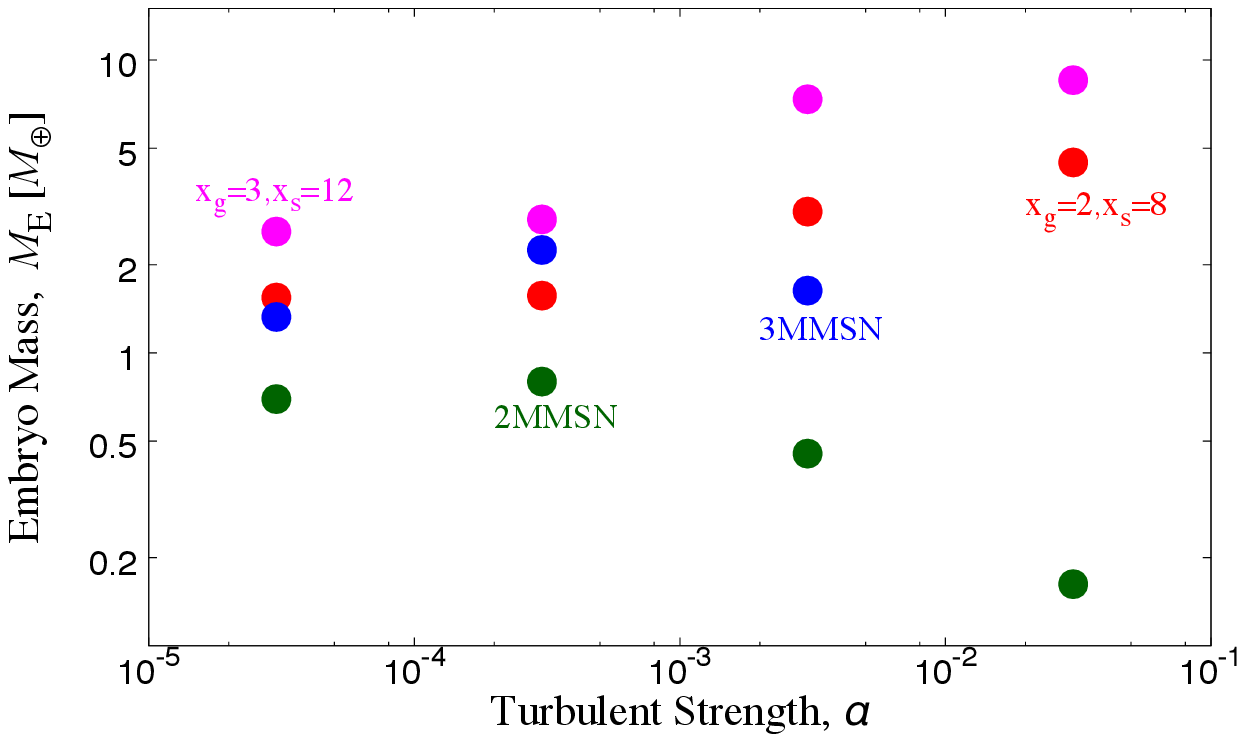}
\plotone{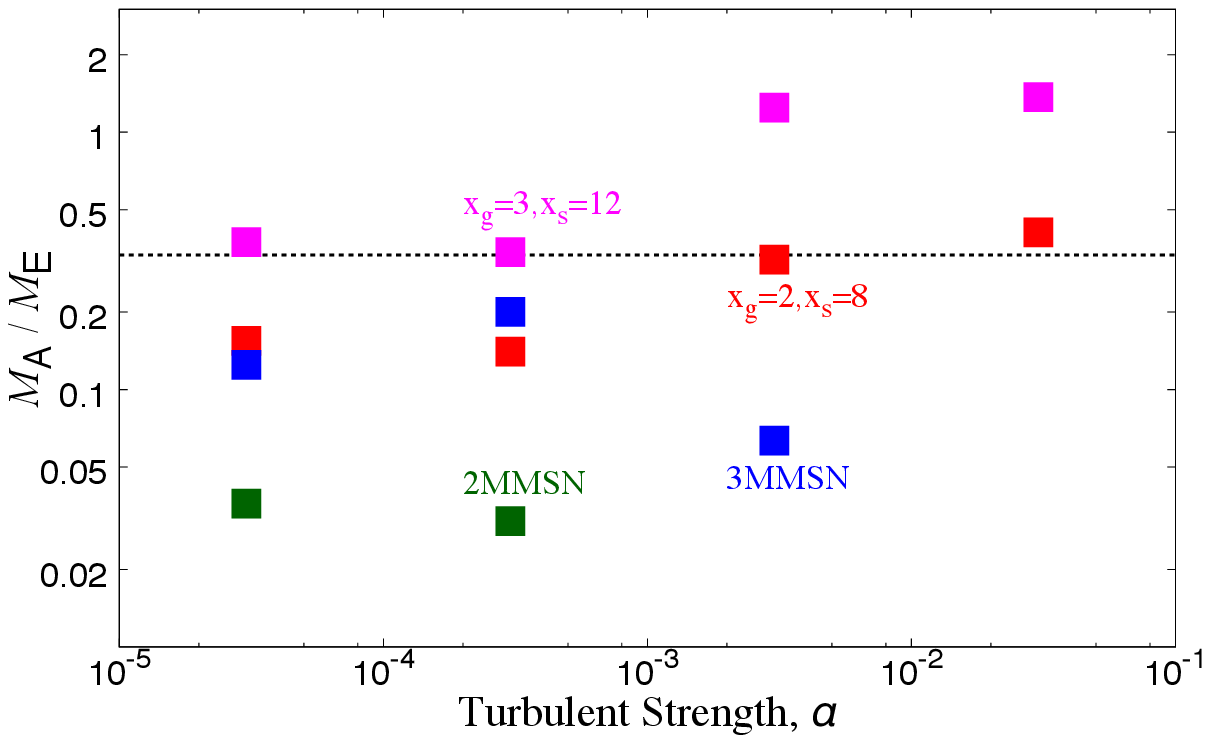}
 \figcaption{
The masses of largest bodies, embryo mass $M_{\rm E}$ (left panel) and 
the mass ratio of atmosphere to core $M_{\rm A} / M_{\rm E}$ (right
 panel) for the
 primordial strength bodies at 4\,Myr at 5.2\,AU in the disks with
 $x_{\rm g} = x_{\rm s} = 2$ (2MMSN; green), 
 $x_{\rm g} = x_{\rm s} = 3$ (3MMSN; blue), 
 $x_{\rm g} = 2$ and $x_{\rm s} = 8$ (red), 
and  $x_{\rm g} = 3$ and $x_{\rm s} = 12$ (magenta). 
Once $M_{\rm A} / M_{\rm E}$ exceeds $1/3$, runaway gas accretion to form gas giants occurs. 
  \label{fig:mlarge_atm_prim} }
\end{figure*}

As discussed in \S~\ref{sc:ccm}, gas giant formation via gas accretion
requires embryos to be larger than $\sim 5 M_\oplus$ prior to
significant gas depletion. 
Fig.\ref{fig:mlarge_atm_prim} shows embryo masses $M_{\rm E}$ at
5.2\,AU at 4\,Myr with primordial strength. 
Massive embryos tend to be formed in the disks with strong turbulence
(large $\alpha$), while too strong turbulence cannot produce massive
embryos because the onset of runaway growth is too late. 
In 3MMSN disks, embryos grow up to 1 -- $2 M_\oplus$, which are too
small to start gas accretion. However, the critical core mass depends on
$\dot M_{\rm E}$. To confirm the possibility of gas giant formation, 
we calculate the masses of static atmospheres around embryos, 
$M_{\rm A}$, based on the analytical model for atmospheric radial density
profile ignoring the gravity of atmospheres \citep{inaba_ikoma03}. The
mass ratio $M_{\rm A}/M_{\rm E}$ is smaller than 0.2 so that runaway gas
accretion does not occurs within 4\,Myr. In the later stage, embryos
grow more massive (see Fig.~\ref{fig:growth_alpha_prim}); 
$M_{\rm A}/M_{\rm E} \ga 1/3$ at $t \approx 5\,$Myr and 6--7Myr 
for $\alpha = 3
\times 10^{-4}$ and $3 \times 10^{-3}$, respectively.
In 2MMSN, the $\alpha$ dependence of $M_{\rm E}$ and $M_{\rm A}$ is
similar to that in 3MMSN, while $M_{\rm E}$ and $M_{\rm A}$ are
smaller. 
Therefore, embryos cannot reach the critical core mass within the disk
lifetime $\la 4\,$Myr in the disks less massive than 3MMSN. 

\begin{figure}[htbp]
 \epsscale{1.} 
\plotone{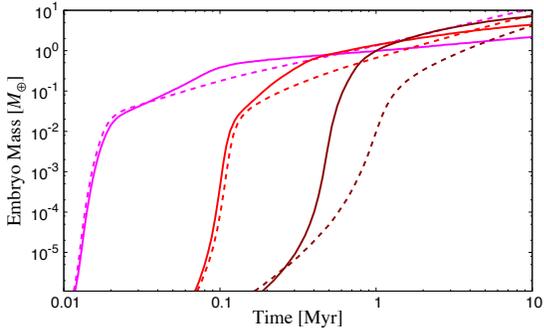}
 \figcaption{
Same as Fig.~\ref{fig:growth_alpha_prim}, but in the disk with $x_{\rm
 g} = 2$ and $x_{\rm s}$ for $\alpha = 3 \times 10^{-4}$ (magenta), $3
 \times 10^{-3}$ (red), and $3 \times 10^{-2}$ (blown). 
  \label{fig:growth_enhanced} }
\end{figure}

The collisional evolution of fluffy dust aggregates overcomes the radial
drift barrier if the dust aggregates that drift
most effectively are controlled in the Stokes gas drag regime, resulting in
planetesimals within $\sim 10$\,AU \citep{okuzumi12}. 
During collisional evolution, the radial drift of dust aggregates
induces a pile-up in the planetesimal forming region so that 
the solid surface density increases by a factor 3 -- 4 \citep{okuzumi12}.  
According to the result, we set $x_{\rm s} / x_{\rm g} = 4$
\footnote{ \revhk{
Although the solid enhancement due to the aggregate growth occurs within
about 
$10$\,AU, the solid surface densities are set to be enhanced
from 4.8 to 26\,AU in the simulations. 
We also conduct the simulation with
the outer disk edge at 9.4\,AU
for $\alpha = 3 \times 10^{-4}$ and $3 \times 10^{-3}$ in the disks with
$x_{\rm g} = 3$ and $x_{\rm s} = 12$ and then  
the masses of embryos at 4\,Myr 
are the same within 20\% for
different outer edges. 
The supply from $\ga 10\,$AU is insignificant because the
late formation of planetary embryos beyond 10\,AU. 
 }
}.
In the solid-enhanced disks with $x_{\rm g} = 3$ and
$x_{\rm s} = 12$, embryo masses at 4\,Myr exceed $5M_\oplus$ 
for $\alpha \ga 3 \times 10^{-4}$ 
so that their atmospheric masses $M_{\rm A}$ are much larger than $M_{\rm E}/3$
(Fig.~\ref{fig:mlarge_atm_prim}). 
For $x_{\rm g} = 2$ and $x_{\rm s} = 8$, $M_{\rm E}$ for $\alpha \la 10^{-3}$ is similar to that
in the case of 3MMSN, while large $M_{\rm E}$ for $\alpha \ga
10^{-3}$ results in $M_{\rm A}/M_{\rm E} \ga 1/3$. 
Fig.~\ref{fig:growth_enhanced} shows the mass evolution of embryos in
the disk with $x_{\rm g} = 2$ and $x_{\rm s} = 8$. The growth is stalled
at $t \ga 0.1$\,Myr for $\alpha = 3 \times 10^{-4}$. 
The embryo growth occurs within $\sim 1$\,Myr even for $\alpha \ga
10^{-3}$, which forms massive cores. 
Therefore, in the solid-enhanced disks, 
runaway gas accretion for
gas giant formation occurs within the disk lifetime $\sim 4\,$Myr. 

For monolith strength, embryo masses are smaller than that for 
primordial strength. In the 3MMSN disks, embryos are too small to start
gas accretion. Even in the 5MMSN disk, $M_{\rm A}/M_{\rm E}$ is smaller than
0.2. Therefore, gas giant formation via core accretion is difficult in
the disk less massive than 5MMSN. 
In the solid enhanced disks with $x_{\rm g} = 3$ and
$x_{\rm s} = 12$, embryo masses reach $\sim 1$ -- $2M_{\oplus}$ for $\alpha \ga
3 \times 10^{-3}$. For $\alpha = 3 \times 10^{-2}$, $M_{\rm A}/M_{\rm E} >
1/3$. 
In the disk with $x_{\rm g} = 5$ and $x_{\rm s} = 20$, $M_{\rm A}/M_{\rm E} >
1/3$ for $\alpha \sim 3 \times 10^{-3}$. 
Therefore, melted bodies may form gas giant planets via core
accretion in more massive disks, comparing to primordial material.

\begin{figure*}[bhtp]
 \epsscale{.45} 
\plotone{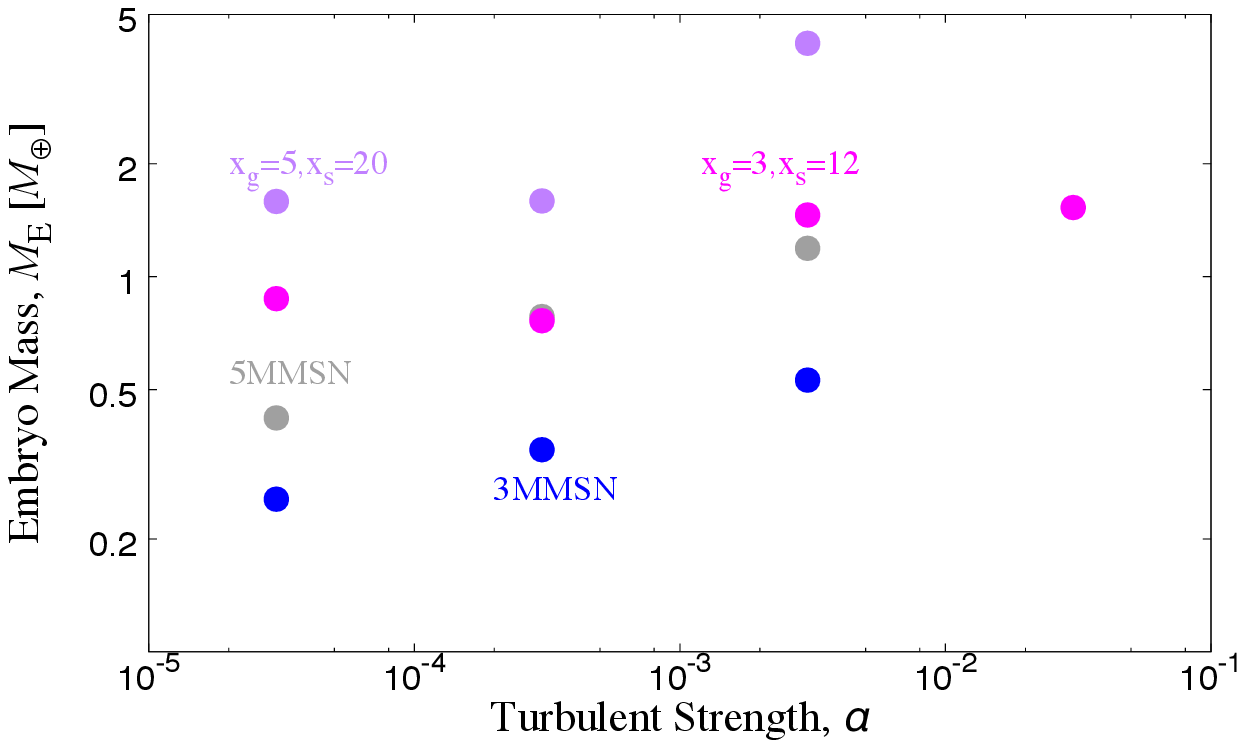}
\plotone{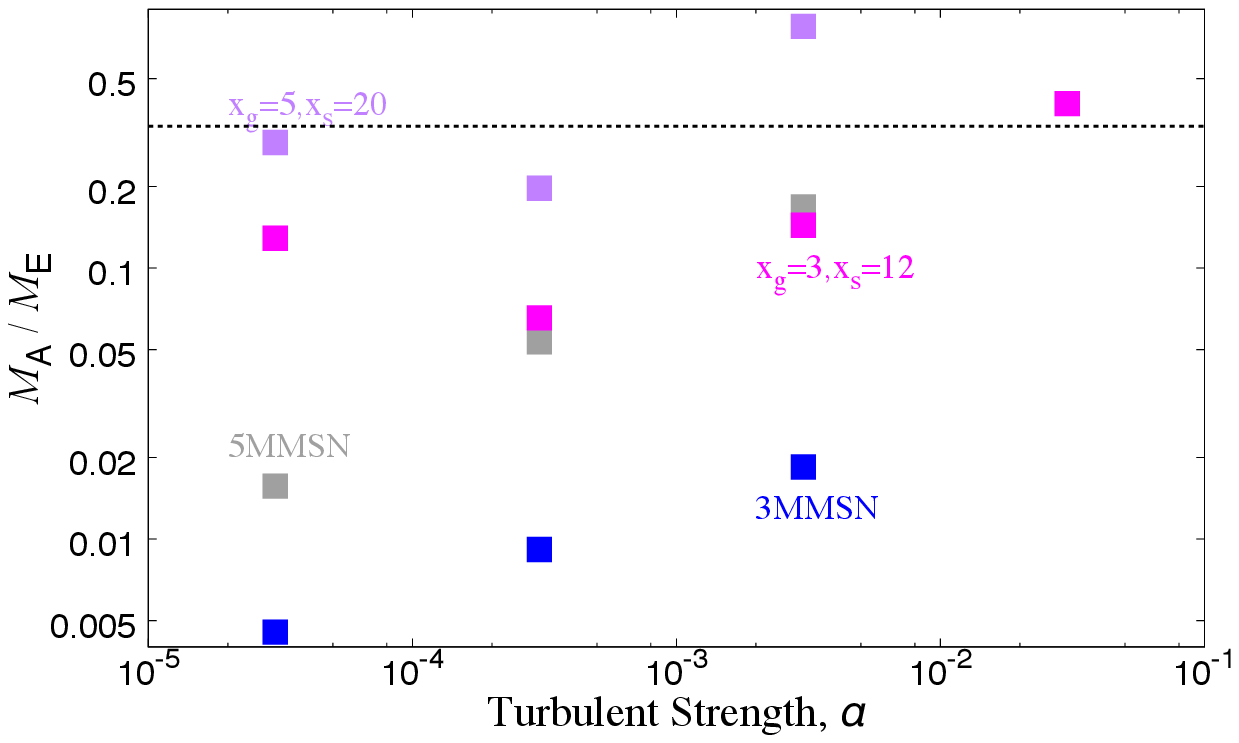}
 \figcaption{
Same as Fig.~\ref{fig:mlarge_atm_prim}, but for monolith strength
in the disks with
 $x_{\rm g} = x_{\rm s} = 3$ (3MMSN; blue), 
 $x_{\rm g} = x_{\rm s} = 5$ (5MMSN; grey), 
 $x_{\rm g} = 3$ and $x_{\rm s} = 12$ (magenta), 
and  $x_{\rm g} = 5$ and $x_{\rm s} = 20$ (purple), 
\label{fig:mlarge_atm_melt} }
\end{figure*}

\section{Discussion}
\label{sc:discussion}

\subsection{Growth Timescale}

Prior to the runaway growth, the mass distribution of bodies is
approximated to be a single mass population \citep{kobayashi16}, so that the collisional timescale at the onset of runaway growth is estimated to
be
\begin{equation}
 \tau_{\rm col} \approx \frac{4 i \rho_{\rm s} r_{\rm rg}}{3 e \Sigma_{\rm
  s} \Omega}. 
\end{equation}
The balance between turbulent stirring and gas drag gives $i/e$ as
$\epsilon$ prior to the onset of runaway growth \citep{kobayashi16}, 
while the gravitational
interaction during runaway growth results in the energy equipartition,
$i/e = 0.5$. The embryo formation timescale via runaway growth is
proportional to $\tau_{\rm col}$ with $i/e = 0.5$ \citep{ormel10a,kobayashi10}. 
The subsequent embryo growth timescale $\tau_{\rm g}$, which depends on the accretion
of planetesimals or small bodies \citep{kobayashi+10,kobayashi11}, 
is simply approximated to be $\tau_{\rm g} \approx 3 \tau_{\rm col}$;
\begin{equation}
 \tau_{\rm g} = 4.5 \left(\frac{x_{\rm s}}{3} \right)^{-1} 
  \left(\frac{r_{\rm rg}}{30\,{\rm km}}\right) 
  \left(\frac{a}{5.2\,{\rm AU}}\right)^3 \,{\rm Myr}.\label{time_rg0}
\end{equation}

The random velocity is determined by the balance between turbulent
stirring and collisional damping prior to runaway growth. Once $v_{\rm
r} \approx 1.5 v_{\rm esc}$, runaway growth occurs. The radius of bodies
at the onset of runaway growth is given by \citep{kobayashi16}
\begin{equation}
 r_{\rm rg} = 32 \left(\frac{\alpha}{3 \times 10^{-3}}\right) 
  \left(\frac{x_{\rm g}}{3}\right)^{2} \left(\frac{x_{\rm
   s}}{3}\right)^{-1}
  \left(\frac{a}{5.2\,{\rm AU}}\right)^{1.5} \,{\rm km}\label{eq:rg_radius}
\end{equation}
From Eqs.~(\ref{time_rg0}) and (\ref{eq:rg_radius}), the timescale of embryo growth depending on turbulence strength is given
by 
\begin{equation}
 \tau_{\rm g} = 4.8 \left(\frac{x_{\rm g}}{x_{\rm s}}\right)^2
\left(\frac{\alpha}{3 \times 10^{-3}}\right) 
\left(\frac{a}{5.2\,{\rm AU}}\right)^{4.5} \,{\rm Myr}.\label{eq:timescale_rg}
\end{equation}
As seen in
Fig.~\ref{fig:mlarge_atm_prim}, growing embryos at $t = 4$\,Myr for
$\alpha = 3 \times 10^{-3}$ are 
less massive than those for $\alpha = 3 \times 10^{-4}$ for $x_{\rm g} = x_{\rm
s}$, while embryos at $t = 4$\,Myr increases with $\alpha$ 
for $x_{\rm s} / x_{\rm g} = 4$. That is explained by the dependence of
$\tau_{\rm g}$ on $x_{\rm s} / x_{\rm g}$;  For $x_{\rm s} / x_{\rm g}
= 4$, $\tau_{\rm g} < 4\,{\rm Myr}$ even for $\alpha \la 3 \times
10^{-2}$, resulting in embryo formation within the
disk lifetime. 

Collisional fragmentation of planetesimals stalls embryo growth so that 
large $r_{\rm rg}$ tends to form embryos
massive enough. The formation of cores with the critical core masses
requires $r_{\rm rg} \ga 10$\,km around 5\,AU for primordial
strength in the disks with $x_{\rm g} \approx 3$. 
On the other hand, large $r_{\rm rg}$ results in a long embryo-growth timescale
(see Eq.~\ref{time_rg0}); $\tau_{\rm g} \ll 4$\,Myr is needed. 
These conditions are not satisfied for 3MMSN, 
while the solid-enhanced disks create massive cores for $\alpha
\ga 10^{-3}$ because of these conditions (see Fig.~\ref{fig:growth_enhanced}). 

\subsection{Planetary Migration and Gas Dispersal}

Planetary embryos migrate due to density waves caused by interaction with
disks; the migration timescale is estimated to be
\citep[e.g.,][]{tanaka}
\begin{eqnarray}
 \tau_{\rm mig} &=& \gamma^{-1} \left(\frac{M_{\rm E}}{M_*} \right)^{-1}
  \left(\frac{\Sigma_{\rm g} a^2}{M_*}\right)^{-1}
  \left(\frac{c_{\rm s}}{v_{\rm K}}\right)^2
  \Omega^{-1},
  \nonumber
\\
 &\approx& 1.2 \gamma^{-1} \left(\frac{M_{\rm E}}{M_\oplus}\right)^{-1}
  \left(\frac{\Sigma_{\rm g}}{450 \, {\rm g \, cm}^{-2}}\right)^{-1}
  \left(\frac{c_{\rm s} / v_{\rm K}}{0.05}\right)^2
\nonumber
\\
&&  \times \left(\frac{a}{5.2\,{\rm AU}}\right)^{-1/2}
  \left(\frac{M_*}{M_\sun}\right)^{3/2} 
\,{\rm Myr}, 
\label{eq:tmig}
\end{eqnarray}
where $\gamma$ is the migration coefficient and the value of
$\Sigma_{\rm g}$ is given from that of at 5.2\,AU in the 3\,MMSN disk. In the isothermal disk,
$\gamma \approx 4$ \citep{tanaka} and the co-rotation torque may
reduce $\gamma \sim 1$ \citep{paardekooper11}. If the migration timescale
is shorter than the time reaching the critical core mass, embryos are
lost due to migration prior to gas giant formation. We here estimate the
migration timescale for embryos
obtained in our simulations using $\gamma = 1$. 

\begin{figure}[htbp]
 \epsscale{1.0} 
\plotone{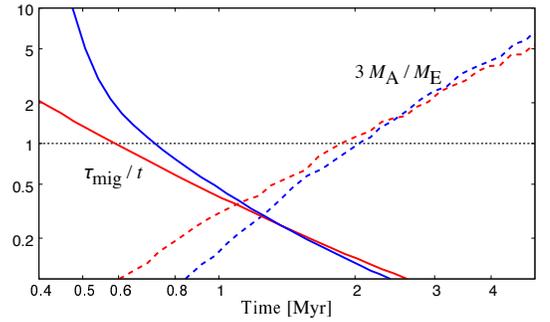}
 \figcaption{
Dimensionless migration timescales $\tau_{\rm mig}/t$ (solid curves) 
calculated from the results of
 simulations with primordial strength in disks with $x{\rm g} = 3$
 and $x_{\rm s} = 12$ for $\alpha = 3 \times 10^{-4}$ (blue) and $3
 \times 10^{-3}$ (red). Dotted curves indicate $3 M_{\rm A}/M_{\rm E}$; 
The onset of runaway gas accretion is estimated from $3 M_{\rm A}/ M_{\rm E} = 1$. 
  \label{fig:tmig} }
\end{figure}

For primordial strength, embryos as large as the critical core mass are
formed within several Myrs in the disk with $x_{\rm s} /x_{\rm g}
\approx 4$ for $\alpha \approx 10^{-3}$ -- $10^{-1}$
(Fig.~\ref{fig:mlarge_atm_prim}). Fig.~\ref{fig:tmig} shows $3 M_{\rm A}
/ M_{\rm E}$ at 5.2\,AU for $\alpha = 3 \times 10^{-4}$ and $3 \times
10^{-3}$; $M_{\rm A}$ becomes $M_{\rm E}/3$ at $t \approx 2$\,Myr and 3
-- 4\,Myr for $\alpha = 3 \times 10^{-4}$ and $3 \times 10^{-3}$,
respectively and embryos then reach the critical core masses.  Using
$M_{\rm E}$ obtained from the simulations and Eq.~(\ref{eq:tmig}), we
calculate $\tau_{\rm mig}/t$. Embryos may grow rather than migration
unless $\tau_{\rm mig} /t \la 1$. Therefore, embryos migrate inward
prior to the formation of embryos with the critical core masses; gas
giant formation is inhibited by migration.

The disk dispersion timescale is required to be comparable to or shorter
than $\tau_{\rm mig}$. \revhk{For $\tau_{\rm gas} = 1$\,Myr, $\tau_{\rm
mig}/t > 1$ is satisfied in the simulations. However the gas surface
density significantly decreases prior to the onset of the gas accretion
of the core ($M_{\rm A} = M_{\rm
E}/3$). Therefore, a more realistic gas dispersal should be taken into
consideration. 
}

\revhk{
For accretion disks with constant $\alpha$ for simplicity, 
the gas surface density is proportional to $\Sigma_{\rm g} \propto
(1+t/\tau_{\rm dep})^{-1.5}$ \citep{lynden-bell}, where $\tau_{\rm dep} =
\alpha^{-1} (v_{\rm K} / c_{\rm s})^2 \Omega^{-1} /3 $ at the disk radius
$r_{\rm cut}$ of
the exponential cutoff for the surface density. 
We estimate 
\begin{eqnarray}
 \tau_{\rm dep} &=& 0.63 \left(\frac{\alpha}{3 \times 10^{-3}}\right)^{-1} 
\left(\frac{c_{\rm s}}{0.1 v_{\rm K}}\right)^{-2}
\left(\frac{r_{\rm cut}}{50\,{\rm AU}}\right)^{3/2} 
\nonumber
\\
&& \times 
 \left(\frac{M_*}{M_\sun}\right)^{-1/2} \,{\rm Myr}. 
\end{eqnarray}
Therefore, $\Sigma_{\rm g}$ may decrease on the timescale of 1\,Myr. 
}

\begin{figure}[htbp]
 \epsscale{1.} 
\plotone{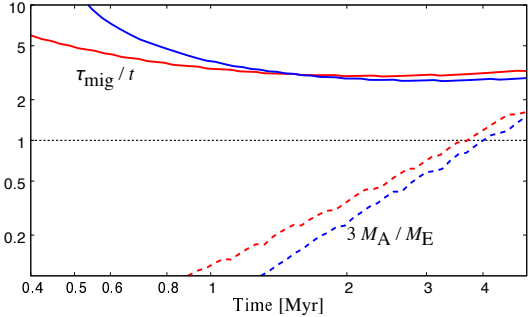}
 \figcaption{
Same as Fig.~\ref{fig:tmig}, but for $\Sigma_{\rm g} = \Sigma_{\rm
 g,0}/(1+t/\tau_{\rm dep})^{3/2}$ with $\tau_{\rm dep} = 10^6$\,yr. 
  \label{fig:tmig_tgas1e6} }
\end{figure}

We perform the simulations assuming $\Sigma_{\rm g} \propto \Sigma_{\rm
g,0}/(1+t/\tau_{\rm dep})^{3/2}$ 
with $\tau_{\rm dep} = 0.5$\,Myr (Fig.~\ref{fig:tmig_tgas1e6}), which mimics to the accretion disk.
Embryos reach the critical core mass at 3\,--\,4\,Myr. 
The early gas depletion weakens the damping of eccentricity and
inclination due to gas drag, and then the accretion of planetesimals on
embryos is suppressed due to large $e$ and $i$. However, small bodies
produced by collisional fragmentation feel the Stokes gas drag, which is
independent of gas density. The early gas depletion does not affect the
accretion of such small bodies significantly. Therefore, embryos exceed the critical
core mass within 4\,Myr. On the other hand, the early gas depletion affects
migration significantly. The gas depletion with $\tau_{\rm dep} \la
1$\,Myr prolongs $\tau_{\rm mig}$, resulting in $\tau_{\rm mig}/t >1$. Therefore, embryos
may start rapid gas accretion prior to migration.

The disk with $x_{\rm g} = 3$ initially has $\sim 17 M_{\rm J}$, where
we set the disk edge at 50\,AU and $M_{\rm J}$ is the mass of Jupiter.
\revhk{The disk mass decreases with $\propto (1+t/\tau_{\rm dep})^{-1/2}$}
\citep{lynden-bell} so that the disk mass becomes $6 M_{\rm J}$ at
$t=4$\,Myr for $\tau_{\rm dep} = 0.5$\,Myr.  The gas accretion of
embryos with the critical core masses occurs not only from around
embryos with the critical core masses but also from the whole disk
during disk evolution \citep{tanigawa07}, so that embryos may acquire an
atmosphere comparable to Jupiter.  Therefore, gas accretion occurs in
such a depleting disk, which saves forming gas giants from type II
migration \citep[e.g.,][]{ida_lin08}.  \revhk{ It should be noted that
$\tau_{\rm dep}$ does not correspond to the disk lifetime inferred from
the infrared observation. In the self-similar solution for accretion
disks \citep{lynden-bell}, the disk evolution timescale prolongs for $t
\gg \tau_{\rm dep}$.  Even for $\tau_{\rm dep} \la 1$\,Myr, the large
amount of gas remaining in several Myrs may be compatible with the disk
lifetime from observations.  }

\section{Summary}
\label{sc:sum}

We investigate planet formation in a turbulent disk. Turbulence
suppresses the runaway growth of planetesimals. Once the random velocity of
planetesimals is comparable to their escape velocity, runaway growth occurs
\citep{kobayashi16}.  The mass-weighted average radius during and after
runaway growth is approximated to be that at
the onset of runaway growth, $r_{\rm rg}$. Embryos formed through
runaway growth become massive through the accretion of planetesimals
with $r \sim r_{\rm rg}$. However, the stirring by massive embryos
induces destructive collisions of planetesimals. Collisional cascade
grinds bodies down to 10 -- 100\,m. Embryos grow through effective
accretion of small bodies, while radial drift reduces small
bodies. Eventually, the embryo growth stalls due to the depletion of
bodies surrounding embryos via destructive collisions and radial
drift. Therefore, the formation and growth of embryos strongly depends
on the collisional properties for $r \sim r_{\rm rg}$, which is
controlled by turbulent strength. 

We have carried out simulations of collisional evolution of bodies with
collisional strengths (Fig.~\ref{fig:qds}) 
for the formation and growth of embryos in
various disks, especially taking into account the stirring by density
fluctuation caused by turbulence (Eq.~\ref{eq:fd}). We find the followings. 
\begin{itemize}
 \item Strong turbulence delays the onset of runaway growth and increases
       $r_{\rm rg}$. Embryos forming within disk lifetimes, $t \approx
       4$\,Myr, tend to be large for high $\alpha$, while 
       the onset of runaway growth is too late to form massive
       embryos within disk lifetimes for $\alpha \ga 10^{-3}$. 
       Cores massive enough are formed from $r_{\rm g} \ga 10\,$km
       corresponding to $\alpha \ga 10^{-3}$. However, the formation
       timescales of such cores are longer than the disk depletion
       time for $\alpha \ga 10^{-3}$. 
 \item Solid-enhanced disks are preferable for the formation of massive
       embryos. Such a local enhancement of solid can occur through
       radial drift of dust aggregates. 
       For weak turbulence
       $\alpha \la 10^{-3}$, embryo masses at $t = 4$Myr are similar
       even in the enhancement of solid. However, embryos may grow
       within the disk lifetime even for strong turbulence $\alpha \ga
       10^{-3}$. Therefore, embryos are as large as the critical core mass
       within disk lifetimes in the gas disk more massive than 2MMSN. 
 \item The embryo growth depends on collisional strength of bodies. We
       investigate collisional evolution for primordial and melted
       bodies using the model described in
       Eqs.~(\ref{eq:qds})--(\ref{eq:betas_porous}) as shown in
       Fig.~\ref{fig:qds}. For weak turbulence disks, embryo growth is
       independent of collisional strength until $M_{\rm E} \approx 0.01
       M_\oplus$. However, once collisional fragmentation of bodies with
       $r\sim r_{\rm rg}$ is effective, the embryo growth strongly
       depends on collisional strength. The low-mass end of collisional
       cascade is important for the efficiency of the accretion of
       bodies with $r = 10$--100\,m. The collisional strength 
       $Q_{\rm D}^* \ga 10^7 \,\rm{erg/g}$ for $r \la 1$\,km, which is satisfied 
       for primordial bodies rather than monolith material, is likely to form a massive core to be a gas
       giant (see Figs.~\ref{fig:mlarge_atm_prim} and
       \ref{fig:mlarge_atm_melt}).  
 \item We have estimated the timescale of planetary migration during embryo
       growth. We have taken into
       account the gas density evolution similar to the accretion disk
       model. For example, 
       embryos grow to the critical core mass for gas giant formation
       prior to migration 
       in the disk initially containing 3MMSN gases for the dispersal
       timescale of 0.5\,Myr 
       (see Fig.~\ref{fig:tmig_tgas1e6}). In spite of the short
       dispersal timescale, the disk mass at the onset of
       rapid gas accretion remains much larger
       than Jupiter masses.  Therefore, moderately massive disks with
       short dispersal timescales are likely to form gas giants without significant migration. 
\end{itemize}

The work is supported by
Grants-in-Aid for Scientific Research (26287101,17K05632,17H01105,17H01103)
from MEXT of Japan and by JSPS Core-to-Core Program ``International Network of Planetary Sciences''. 

\appendix 

\section{planetary atmosphere}
\label{ap:atom_dens}

The atmosphere structure is governed by \citep[e.g.,][]{inaba_ikoma03}
\begin{eqnarray}
 \frac{dP}{dR} &=& - \frac{GM_{\rm E} \rho_{\rm a}}{R^2},\label{eq:dp_dr} \\
\frac{dT}{dR} &=& 
\left\{ 
\begin{array}{lll}
\displaystyle
&- \frac{3}{4 \pi \sigma_{\rm SB}} \frac{\kappa L_{\rm e}
P}{T^4} \quad &{\rm if} \quad \frac{3 \kappa L_{\rm e}
P}{4 \pi \sigma_{\rm SB} T^4} < 1 - \frac{1}{\Gamma_2}, \\
\displaystyle
&- \left(1-\frac{1}{\Gamma_2}\right) \frac{G M \rho_{\rm a}}{P R^2 T} \quad &{\rm
 otherwise}, 
\end{array}
\right.\label{eq:dT_dr}
\end{eqnarray}
where $P$ is the pressure, $T$ is the temperature, and $\Gamma_2$ is the
second adiabatic exponent. 
In this study, we assume the atmospheric mass $M_{\rm A}$ is much
smaller than $M_{\rm E}$ . However, if $M_{\rm A}/M_{\rm E} \ga 1/3$,
Eq.~(\ref{eq:dp_dr}) is invalid and then the
rapid gas accretion forms gas giants \citep[e.g.][]{mizuno80}. 

In Eq.~(\ref{eq:dT_dr}), the first term on
the right hand side is given by the radiative energy transfer, while the second
term is determined by convective transfer. 
Here, we derive a solution
using the radiative term, although we take into account the both terms
for simulations below \S~\ref{Sc:model}. Dividing Eq.~(\ref{eq:dT_dr})
by Eq.~(\ref{eq:dp_dr}) and integrating over $P$, we have 
\begin{equation}
 \rho_{\rm a} = \frac{16 \pi \sigma_{\rm SB} G M_{\rm E} T^3 \mu m_{\rm H}}{3 k
  \kappa L_{\rm e}}.\label{eq:rho_mae} 
\end{equation}
Inserting Eq.~(\ref{eq:rho_mae}) into Eq.~(\ref{eq:dT_dr}) and
integrating it, we then obtain
\begin{equation}
 T = \frac{GM_{\rm E} \mu m_{\rm H}}{4 k R}.\label{170423_8Mar18} 
\end{equation}
Substitution of Eq.~(\ref{170423_8Mar18}) into Eq.~(\ref{eq:rho_mae}) gives
Eq.~(\ref{eq:rho_a}).

\section{collisional outcome modeling}
\label{app_collisional_modeling}

Collisional outcome from the collision between bodies with $m_1$ and $m_2$ is expressed by $m_{\rm e}$ and $\Psi(m,m_1,m_2)$,
which are determined as follows. 
\begin{equation}
 m_{\rm e} = \frac{\phi}{1+\phi} (m_1+m_2),  
\end{equation}
where $\phi$ is the dimensionless impact energy. Using the impact
velocity $v_{\rm imp}$ and $Q_{\rm D}^*$, which is the specific impact
energy needed for ejection of half bodies, 
$\phi = m_1 m_2 v_{\rm imp} / 2 (m_1 + m_2)^2 Q_{\rm D}^* $. 
\begin{equation}
 (m_1+m_2) \Psi(m,m_1,m_2) = 
  \left\{
   \begin{array}{ll}
    \displaystyle
     m_{e} & (m>m_{\rm L}),
     \\
    \displaystyle
     m_{e} \left(\frac{m}{m_{\rm L}}\right)^{2-b} & (m \leq m_{\rm L}),
   \end{array}
  \right.\label{eq:f_eject}
\end{equation}
where $\phi$ is the dimensionless impact energy and 
$m_{\rm L}$ is the largest mass of fragments produced by a single
collision between bodies with $m_1$ and $m_2$, given by 
\begin{equation}
 m_{\rm L} = \frac{\epsilon_{\rm L} }{1+\phi}  m_{\rm e} =
  \frac{\epsilon_{\rm L} 
   \phi}{(1+\phi)^2} (m_1+m_2),\label{eq:mlarge} 
\end{equation}
$b$ and $\epsilon_{\rm L}$ are constants. We set $b = 5/3$ and
$\epsilon_{\rm L} =0.2$. 
The timescale of collisional cascades is insensitive to values of
$b$ and $\epsilon_{\rm L}$ \citep{kobayashi10}.

\end{document}